\documentclass[twocolumn,aps,prc,superscriptaddress]{revtex4-2}

\usepackage{amssymb}
\usepackage{amsmath}
\usepackage{graphicx}
\usepackage[normalem]{ulem}
\usepackage{multirow}
\usepackage{appendix}
\usepackage{CJK}
\usepackage[usenames]{color}
\usepackage{bm}
\usepackage{booktabs}
\usepackage{threeparttable}
\usepackage[colorlinks,linkcolor=blue,urlcolor=blue,citecolor=blue]{hyperref}
\makeatletter
\def\UrlAlphabet{%
	\do\a\do\b\do\c\do\d\do\e\do\f\do\g\do\h\do\i\do\j%
	\do\k\do\l\do\m\do\n\do\o\do\p\do\q\do\r\do\s\do\t%
	\do\u\do\v\do\w\do\x\do\y\do\z\do\A\do\B\do\C\do\D%
	\do\E\do\F\do\G\do\H\do\I\do\J\do\K\do\L\do\M\do\N%
	\do\O\do\P\do\Q\do\R\do\S\do\T\do\U\do\V\do\W\do\X%
	\do\Y\do\Z}
\def\UrlDigits{\do\1\do\2\do\3\do\4\do\5\do\6\do\7\do\8\do\9\do\0}
\g@addto@macro{\UrlBreaks}{\UrlOrds}
\g@addto@macro{\UrlBreaks}{\UrlAlphabet}
\g@addto@macro{\UrlBreaks}{\UrlDigits}
\makeatother
\renewcommand{\sout}{\bgroup \color{red} \ULdepth=-.5ex \ULset}

\newcommand{\hyt}{$^3_\Lambda \text{H}$}

\begin{document}
\preprint{APS/123-QED}

\title{Anisotropic Flow of Light (Anti-)(hyper-)nuclei in Pb+Pb Collision at \texorpdfstring{$\sqrt{s_{NN}}=5.36$}{} TeV}
\author{Fu Ma}
\affiliation{Key Laboratory of Nuclear Physics and Ion-beam Application~(MOE), Institute of Modern Physics, Fudan University, Shanghai $200433$, China} 
\affiliation{Shanghai Research Center for Theoretical Nuclear Physics, NSFC and Fudan University, Shanghai 200438, China}

\author{Zheng-Qing Wang}
\affiliation{Key Laboratory of Nuclear Physics and Ion-beam Application~(MOE), Institute of Modern Physics, Fudan University, Shanghai $200433$, China} 
\affiliation{Shanghai Research Center for Theoretical Nuclear Physics, NSFC and Fudan University, Shanghai 200438, China}

\author{Xiong-Hong He}
	\email{hexh@impcas.ac.cn}
    \affiliation{School of Nuclear Science and Technology, University of Chinese Academy of Sciences, Beijing 100049, China}
    \affiliation{Heavy lon Science and Technology Key Laboratory, Institute of Modern Physics,
 Chinese Academy of Sciences, Lanzhou 730000, China}

	\author{Che Ming Ko}
	\email{ko@comp.tamu.edu}
	\affiliation{Cyclotron Institute and Department of Physics and Astronomy, Texas A\&M University, College Station, Texas 77843, USA}
\author{Yu-Gang Ma} 
\email{mayugang@fudan.edu.cn} 
\affiliation{Key Laboratory of Nuclear Physics and Ion-beam Application~(MOE), Institute of Modern Physics, Fudan University, Shanghai $200433$, China} 
\affiliation{Shanghai Research Center for Theoretical Nuclear Physics, NSFC and Fudan University, Shanghai 200438, China}
\affiliation{School of Physics, East China Normal University, Shanghai 200241, China}

 \author{Qi-Ye Shou}
 \email{shouqiye@fudan.edu.cn}
\affiliation{Key Laboratory of Nuclear Physics and Ion-beam Application~(MOE), Institute of Modern Physics, Fudan University, Shanghai $200433$, China} 
\affiliation{Shanghai Research Center for Theoretical Nuclear Physics, NSFC and Fudan University, Shanghai 200438, China}  

	\author{Kai-Jia Sun}
	\email{kjsun@fudan.edu.cn}
\affiliation{Key Laboratory of Nuclear Physics and Ion-beam Application~(MOE), Institute of Modern Physics, Fudan University, Shanghai $200433$, China} 
	\affiliation{Shanghai Research Center for Theoretical Nuclear Physics, NSFC and Fudan University, Shanghai 200438, China}

\author{Wenbin Zhao}
\email{ wenbinzhao@ccnu.edu.cn} 
\affiliation{ Institute of Particle Physics and Key Laboratory of Quark and Lepton Physics (MOE)}
\affiliation{Central China Normal University, Wuhan, 430079, Hubei, China} 

\author{Wen-Hao Zhou}
\email{zhouwenhao@xaau.edu.cn}
\affiliation{Faculty of Science, Xihang University, Xi'an, 710077, China}

\begin{abstract}
    Using the coalescence model with nucleon phase-space distributions generated  by the hybrid MUSIC framework, we study the elliptic flow ($v_2$) and triangular flow ($v_3$) of (anti-)protons, (anti-)deuterons, (anti-)$^3\mathrm{He}$, and ${^3_\Lambda\mathrm{H}}$ in Pb+Pb collisions at $\sqrt{s_{NN}} = 5.36$ TeV. We find that the simple $v_2$ scaling with the number of constituent nucleons $A$ breaks down at high transverse momentum $p_T/A > 1.5$ GeV/$c$, while an improved scaling  relation  holds well up to $p_T/A \approx 3$ GeV/$c$. In contrast, $v_3$ exhibits similar behavior under both scaling prescriptions, with no significant difference. We also make predictions for $v_2$ and $v_3$ of the hypertriton and find these flows are insensitive to the Lambda-deuteron ($\Lambda-d$) distance inside the hypertriton. Our results are compared with preliminary experimental measurements by the ALICE Collaboration and offer insight into the production mechanisms of light (anti-)(hyper-)nuclei in high-energy heavy-ion collisions. 
\end{abstract}
\maketitle

\section{INTRODUCTION}
Light nuclei, such as  the deuteron ($d$), helium-3 ($^3$He) and helium-4 ($^4$He),    as well as their antiparticles, have been measured in high-energy heavy-ion collisions at the Relativistic Heavy Ion Collider (RHIC) and the Large Hadron Collider (LHC)~\cite{ALICE:2020foi,STAR:2010gyg,STAR:2011eej,ALICE:2015rey,STAR:2019wjm,ALICE:2015wav,STAR:2009tro,Chen:2024,Shou:2024,Jia:2024}. These bound-state nuclei have small binding energies and relatively large sizes compared with the temperature and  spatial extent of the fireball produced in high-energy nuclear collisions. Their production mechanism remains a topic of active  debate \cite{Oh:2009gx,Feckova:2016kjx,Mrowczynski:2016xqm,Bazak:2018hgl,Oliinychenko:2018ugs,Bellini:2018epz,Bugaev:2018klr}. In the statistical hadronization model (SHM), light nuclei are produced at a common chemical freeze-out temperature  following hadronization of the quark-gluon plasma (QGP)~\cite{Vovchenko:2019kes,Andronic:2010qu,Andronic:2016nof}. In contrast,   in the nucleon coalescence model~\cite{Scheibl:1998tk,Sato:1981ez,Sun:2018mqq}, light nuclei  are  formed at the kinetic freeze-out stage, when the temperature and density of the hadronic matter are much lower.  More recently, a kinetic model has been developed that accounts for the dissociation and regeneration of light nuclei through pion-catalyzed reactions during the expansion of the hadronic matter~\cite{Sun:2022xjr,Wang:2023gta,Oh:2007vf,Oliinychenko:2018ugs,Danielewicz:1991dh}. 

In heavy-ion collisions, the large pressure gradients generated in the early stages convert the spatial anisotropies of the initial collision geometry into anisotropic momentum distributions of the produced particles. These azimuthal distributions are conventionally characterized by a Fourier series~\cite{Poskanzer:1998yz}:  
\begin{equation}
\frac{dN}{d\phi} \propto 1 + \sum_{n=1}^{\infty} 2v_n \cos[n(\phi - \psi_n)],
\label{eq_phi}
\end{equation}
where $\phi$ is the azimuthal angle of the particle in momentum space, $\psi_n$ is the azimuthal angle of the $n$-th order symmetry plane, and $v_n$ is the $n$-th order  anisotropic flow coefficient, with $v_2$ and $v_3$ referred to as elliptic and triangular flow, respectively. Significant anisotropic flow signals have been observed not only in heavy-ion collisions but also in smaller collision systems, including  proton-proton ($pp$)~\cite{CMS:2010ifv,ATLAS:2015hzw,CMS:2015fgy,CMS:2016fnw} and proton-nucleus~\cite{CMS:2012qk,ALICE:2012eyl,ATLAS:2012cix,CMS:2013jlh,ATLAS:2014qaj,CMS:2014und,CMS:2015yux,LHCb:2015coe} collisions.  Anisotropic flow provides a powerful tool for probing particle production mechanisms through coalescence scaling relations.  In the coalescence picture, nucleons (or quarks)    that are close  in phase space have a higher probability to   combine to form  light nuclei (or hadrons), and the resulting composite particles inherit the flow of their constituents. Specifically, if  $A$ nucleons with similar momenta coalesce into a nucleus, the azimuthal distribution  of nucleus follows~\cite{Kolb:2004gi,Molnar:2003ff,Yan:2006bx,Ma:2007}: 
\begin{equation}\label{eq:ImprovedScaling}
\frac{d N_A}{d\phi} \propto  \left(\frac{d N_p}{d\phi}\right)^{A}.
\end{equation}
In the limit of small anisotropies ( $v_n\ll 1$ ), this yields the approximate scaling relation 
 \begin{equation}
v_n(p_T) \approx A \, v_n^{(p)}(p_T / A), 
\label{eq:SimpleScaling}
\end{equation}
known as the number-of-constituent nucleon    (NCN) scaling, which  is a key  prediction of the nucleon  coalescence model for light nuclei production as first proposed by Yan and Ma {\it et al.} ~ \cite{Yan:2006bx,Ma:2007} and confirmed by STAR and BM@N experimental data \cite{STAR1,STAR2,BMN}. The earlier analogous relation, the number-of-constituent-quark (NCQ)  scaling of $v_2$,  has been experimentally confirmed for hadrons  at both RHIC~\cite{STAR:2003wqp, PHENIX:2006dpn} and the LHC~\cite{ALICE:2014wao,ALICE:2018yph,CMS:2014und}, as well as in $p$+Pb collisions by the CMS  
 and ALICE experiments~\cite{CMS:2014und,CMS:2018loe,ALICE:2024vzv} . These observations validate the quark  coalescence mechanism for hadron production, particularly at intermediate $p_T$~\cite{Zhao:2020wcd,Zhao:2021vmu}. Extending this approach to light nuclei, a systematic study of the NCN  scaling behavior of  their anisotropic flow coefficients relative to those of their constituent nucleons can provide direct insight into the production mechanisms of light nuclei \cite{Yan:2006bx,Ma:2007,Han:2011iy,Wang:2019,Hillmann:2019wlt,Du:2023ype}.

Recently, the ALICE collaboration has measured the   $v_2$  of deuterons  and $^3$He in Pb+Pb collisions at $\sqrt{s_{NN}}=5.02$ TeV~\cite{ALICE:2018yph, ALICE:2020chv, ALICE:2019ikx}. The   measured  (anti-)$^3$He $v_2$ across all centrality bins  is found to lie  between the predictions     of  the Blast-Wave model and  those of  a simple coalescence approach. A more realistic coalescence model,   integrating hydrodynamics evolution followed with the hadronic afterburner UrQMD, provides  an improved   description of the data in the transverse momentum range    $2 < p_T < 6$~ GeV/$c$ for   the $0$–$20\%$ and $20$–$40\%$ centrality classes~\cite{Yin:2017qhg, Zhu:2017zlb, Zhao:2018lyf}.  

In the present study, we investigate the elliptic flow    and triangular flow    of (anti-)protons, (anti-)deuterons , and (anti-)$^3$He in Pb+Pb collisions at $\sqrt{s_{NN}} = 5.36$TeV,     employing the nucleon  coalescence model with nucleon phase-space distributions generated   by  the hybrid MUSIC   framework. The  NCN scaling  of both $v_2$ and $v_3$ for deuterons and  $^3$He is examined over the range  $p_T/A\approx 3$~GeV/$c$ (A=1 for $p$, A=2 for $d$ and A = 3 for $^3$He). We find that the simple scaling relation  with the number of constituent nucleons ($A$) breaks down for $v_2$  at high transverse momentum ($p_T/A > 1.5$ GeV/$c$), whereas  an improved scaling relation  remains valid up to $p_T/A \approx 3$ GeV/$c$. In addition, we  study the elliptic   and triangular flows of  the  (anti-)hypertriton ($^3_\Lambda$H) in    the same collision system   and investigate the   sensitivity  of its anisotropic flow    coefficients to  the spatial separation between the $\Lambda$ hyperon and the deuteron core  within  the $^3_\Lambda$H  wave function.

\section{\label{sec:level2}Methods}
In this study, we employ the hybrid MUSIC+UrQMD+COAL framework to generate the phase-space distributions of nucleons at kinetic freeze-out for coalescence model calculations of light nuclei  production  in Pb+Pb collisions at  $\sqrt{s_{NN}} = 5.36~\mathrm{TeV}$. The quark-gluon plasma (QGP) formed in these collisions  is evolved using the (2+1)-dimensional viscous hydrodynamic model MUSIC~\cite{Paquet:2015lta,Shen:2017bsr,Shen:2020mgh}, initialized with the IP-Glasma model~\cite{Schenke:2012wb,Schenke:2012hg}, which incorporates event-by-event fluctuations in the initial geometry  and gluon saturation within the Color Glass Condensate (CGC). The  equation of state employed is NEOS-BQS~\cite{Monnai:2019hkn}, which implements a crossover transition at finite baryon density. Hadron production is realized on a constant energy-density hypersurface via the Cooper-Frye formula~\cite{Cooper:1974mv}, and subsequent  hadronic rescatterings and resonance decays are modeled using the Ultra-relativistic Quantum Molecular Dynamics (UrQMD) transport framework~\cite{Bass:1998ca}. The resulting freeze-out phase-space distributions of nucleons and $\Lambda$ hyperons  are then passed to the coalescence model~\cite{Chen:2003qj,Chen:2003ava,Mattiello:1996gq}  to describe the production of  light (hyper-)nuclei. The parameters in the hydrodynamic simulations are taken from Ref.~\cite{Mantysaari:2024uwn}.

In the nucleon coalescence model, the invariant momentum spectra of  a light  nucleus is  determined by the overlap between  its Wigner function and the phase-space   distributions $f(\bm{r}_i,\bm{p}_i,t)$ of  kinetically frozen-out nucleons and hyperons:
 \begin{eqnarray}
&&E_A\frac{d^{3}{N_{A}}}{d\bm{P}^{3}_{A}}=g_{c}E_A\int\prod^{A}_{i=1}\left(p^{\mu}_{i}d^{3}\sigma_{i\mu}\frac{d^3\bm{p}^i}{E_i}f(\bm{r}_{i},\bm{p}_{i},t)\right)\notag\\
&&\hspace{1.8cm}\times W_A(\bm{r}^{\prime}_{1},\bm{r}^{\prime}_{2},\cdots,\bm{r}^{\prime}_{A},\bm{p}^{\prime}_{1},\bm{p}^{\prime}_{2},\cdots,\bm{p}^{\prime}_{A};t^{\prime})\notag\\
&&\hspace{1.8cm}\times\delta^{(3)}\left(\bm{P_A}-\sum^{A}_{i=1}\bm{p}_{i}\right),\label{eq1}
\end{eqnarray}
 where $g_c = (2J_A + 1)/[\prod_{i=1}^{A}(2J_i + 1)]$ is the statistical factor for $A$ nucleons with  individual spins $J_i$ to form a nucleus with total angular momentum $J_A$.  The  coordinate $\bm{r}_i$ and momentum $\bm{p}_i$ of the $i$-th nucleon or hyperon in the distribution function $f(\bm{r}_i, \bm{p}_i, t)$ are defined on and integrated over the freeze-out hypersurface, while the primed coordinates $\bm{r}'_i$ and momenta $\bm{p}'_i$ entering the Wigner function of the produced nucleus are obtained  via a Lorentz transformation from the laboratory frame to the rest frame of the nucleus.

Following  Ref.~\cite{Chen:2003ava}, we use  products of harmonic oscillator wave functions for nuclei and obtain their Wigner functions in Eq.(\ref{eq1}) from the Wigner transformation. For the deuteron, the resulting Wigner function is
\begin{equation}
	W_{d}(\bm{r},\bm{p})=8g_d\exp\left(-\frac{\bm{r}^2}{\sigma^2_d}-\sigma^2_d
	\bm{p}^2\right),\label{eq2}
\end{equation}
where $g_d = 3/4$ is the statistical factor for two spin-1/2 nucleons to form a spin-1 deuteron. With  $m_i$  denoting the mass of nucleon $i$, the relative coordinate $\bm{r}$ and relative momentum $\bm{p}$ are defined as
\begin{equation}
    \begin{aligned}
    \bm{r} =
	\frac{1}{\sqrt {2}}(\bm{r}_1 - \bm{r}_2) \\
    \bm{p} = \sqrt
	{2}\frac{m_2\bm{p}_1-m_1\bm{p}_2}{m_1+m_2} 
    \end{aligned},\label{eq3}
\end{equation}
where the  size parameter $\sigma_d$ is related to the root-mean-square radius $r_d \approx 1.96$ fm of the deuteron~\cite{Sun:2017ooe, Ropke:2008qk} by $\sigma_{d}=\sqrt{4/3}r_d=2.26$ fm.

Similarly, the Wigner function of helium-3 (hypertriton) is given by
\begin{equation}
    \begin{aligned}
        W_{3}(\bm{r},\bm{p})=& 8^{2}{g_3}\exp\left(
            -\frac{\bm{r}^2}{\sigma^2_{r}}-\frac{\bm{\lambda}^2}{\sigma^2_{\lambda}}
              - \sigma^2_{r} \bm{p}_{r}^2 - \sigma^2_{\lambda}
    \bm{p}_{\lambda}^2 \right),\label{eq4}
    \end{aligned}
\end{equation}
where $\bm{r}$ and $\bm{p}_r$ are defined in the same way as in Eq.~(\ref{eq3}), and the relative coordinate $\bm{\lambda}$ and momentum $\bm{p}_{\lambda}$ are defined as 
\begin{equation}
	\begin{aligned}
        \bm{\lambda}          & =\sqrt\frac{2}{3}\left(\frac{m_1\bm{r}_1+m_2\bm{r}_2}{m_1+m_2}-\bm{r}_3\right),             \\
		\bm{p}_{\lambda} & =\sqrt\frac{3}{2}\frac{m_3(\bm{p}_1+\bm{p}_2)-(m_1+m_2)\bm{p}_3}{m_1+m_2+m_3},
	\end{aligned}\label{eq5}
\end{equation}
\begin{table}
	\caption{\label{tab:table1}
        Statistical factors ($g$) and width parameters ($\sigma_{r}$, $\sigma_{\lambda}$) for the deuteron, $^3$He and $^3_\Lambda$H.
    }
	\begin{ruledtabular}
		\begin{tabular}{cccc}
			Nucleus       & $g$ & $\sigma_{r}$ (fm) & $\sigma_{\lambda}$ (fm) \\
			\hline
			$p+n\rightarrow d$             & 3/4 & 2.26 & /   \\
			$p+p+n\rightarrow ^3$He        & 1/4 & 1.76 & 1.76 \\
			$p+n+\Lambda\rightarrow ^3_\Lambda$H & 1/4 & 2.26 & 5.45/6.52/7.96 \\
		\end{tabular}
	\end{ruledtabular}
\end{table}

The width parameters $\sigma_r$ and $\sigma_{\lambda}$ in Eq.(\ref{eq4}) are determined by the root-mean-square radii of $^3$He and $^3_\Lambda$H as in the case of the deuteron. For the $^3$He, which has a radius of $r_{\rm ^3He}=1.76$ fm~\cite{Ropke:2008qk}, we take $\sigma_r=\sigma_\lambda=r_{\rm ^3He}=1.76$ fm as in Ref.~\cite{Sun:2017ooe}.  

For the lightest known hypernucleus, the hypertriton ($^3_\Lambda$H), and its  antiparticle,  the anti-hypertriton  ($_{\bar{\Lambda}}^3 \overline{\text{H}}$), which was first detected in 2010 by the STAR Collaboration at RHIC~\cite{STAR:2010gyg}, they are  known to have a lifetime close to that of a free $\Lambda$ and a very small $\Lambda$ separation energy of $B_\Lambda=0.17 \pm 0.06$~MeV~\cite{Chen:2023mel,STAR:2021orx,ALICE:2022sco}. This results  in a structure reminiscent of a halo nucleus, in which the weakly bound $\Lambda$ is spatially separated by a large distance from a deuteron-like nucleon pair~\cite{Nemura:1999pv}.  Accordingly, we adopt $\sigma_r=2.26$~fm, the same value as for the deuteron,  while considering three different values of $\sigma_{\lambda}$ corresponding to different root-mean-square $\Lambda-d$ separation distance $r_{\Lambda d}$. Using the relation $\sigma_{\lambda}=\frac{2}{3}r_{\Lambda d}$  together with the connection between $r_{\Lambda d}$ and the $\Lambda$ separation energy $B_\Lambda$ in $^3_\Lambda$H, we obtain the approximate expression $\sigma_{\lambda} \approx (2.15(B_\Lambda/\mathrm{MeV})^{-1/2} + 1.23)$~fm. Based on $B_\Lambda$ values measured by the STAR and ALICE collaborations, we use $\sigma_{\lambda} = 5.45~\mathrm{fm}, 6.52~\mathrm{fm}$, and $7.96~\mathrm{fm}$ for the hypertriton in this study~\cite{Liu:2024ygk}.
These size parameters and the statistical factors for  the nuclei considered in our study are summarized in Table~\ref{tab:table1}.

We note that, in the coalescence model for hypernuclei production, the halo structure of hypertriton, characterized by a large $\Lambda-d$ separation distance of approximately $10$ fm, leads not only to a suppression of the \hyt ~yield~\cite{Sun:2018mqq,Vovchenko:2018fiy} but also to a softening of its transverse-momentum ($p_T$) spectrum, with a weak centrality dependence~\cite{Liu:2024ygk}.

\par

\section{\label{sec:level3}Results} 
\begin{figure}[!h]
    \centering
    \includegraphics[width=0.48\textwidth]{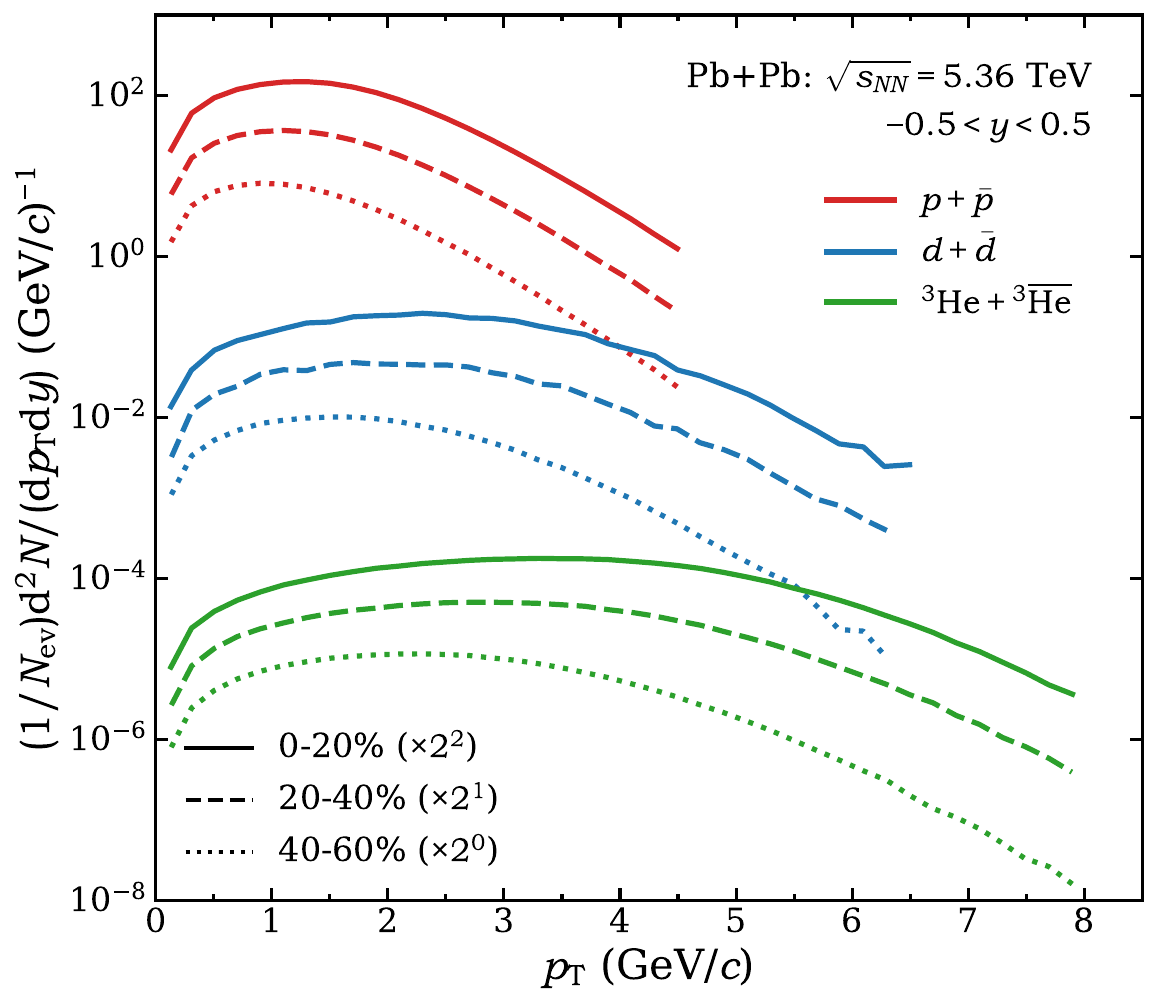}
    \caption{\label{fig:fig1}
    Invariant transverse momentum ($p_T$) spectra of protons ($p+\bar{p}$), deuterons ($d+\bar{d}$), and helium-3 ($^{3}\mathrm{He}+^{3}\overline{\mathrm{He}}$) at midrapidity ($-0.5 < y < 0.5$) in Pb+Pb collisions at $\sqrt{s_{NN}} = 5.36$ TeV. The theoretical predictions from the hybrid dynamical approach coupled with the nucleon coalescence model are shown for three centrality classes (0--20\%, 20--40\%, and 40--60\%).  For  clarity within a single panel, the spectra for different centralities are scaled by factors of $2^n$ ($n=2, 1, 0$).
}
\end{figure}
 \begin{figure*}[!t]
    \centering
    \includegraphics[width=0.95\textwidth]{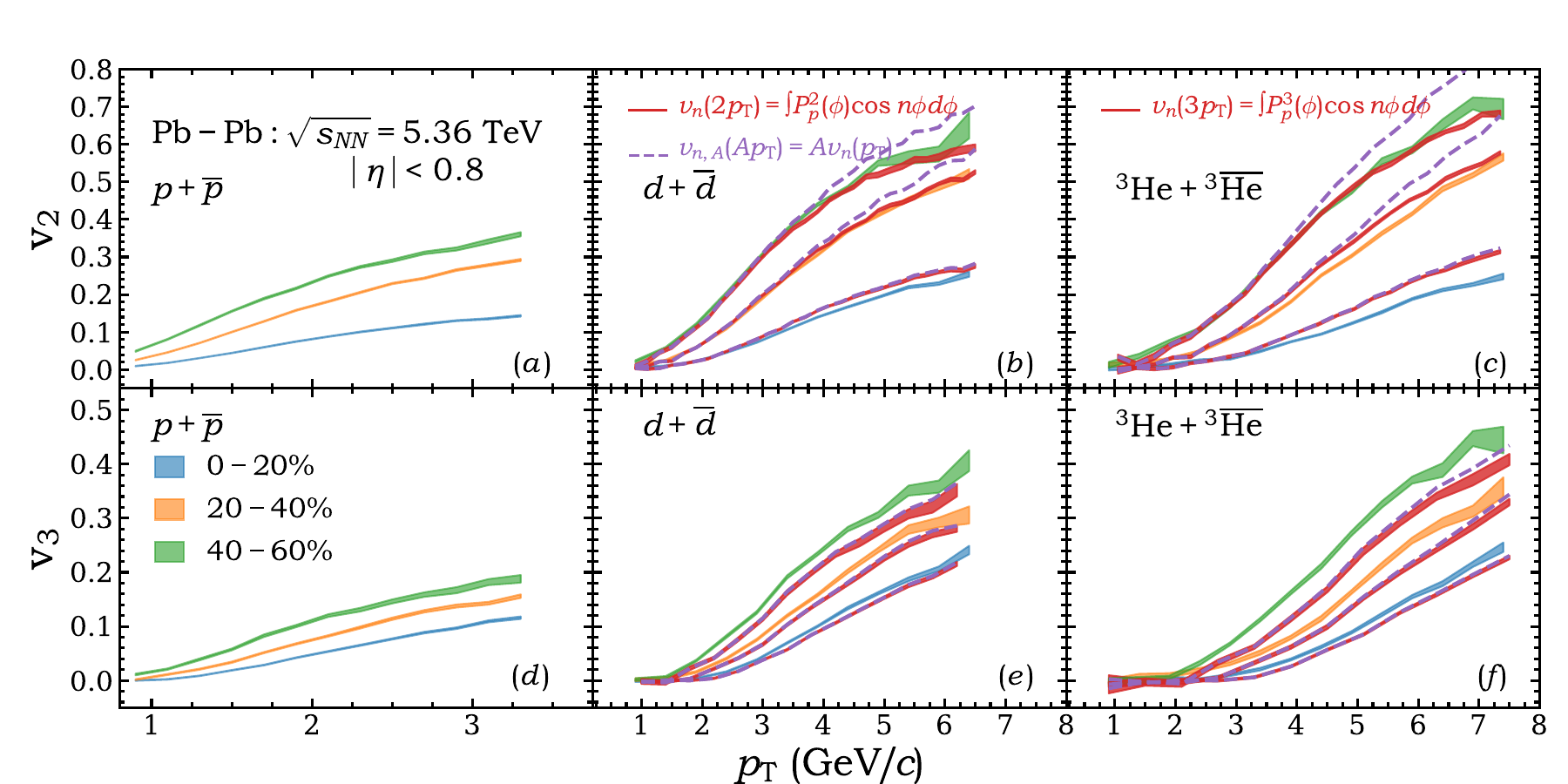}
    \caption{\label{fig:fig2}
   Panels (a)-(f)  show $v_2(p_T)$ and $v_3(p_T)$ for Pb+Pb collisions  at $\sqrt{s_{NN}}=5.36$ TeV in the 0-20\%, 20-40\% and 40-60\% centrality classes for different particles:  (anti-)proton (a, d), (anti-)deuteron (b, e ) and (anti-)Helium-3 (c, f). The solid lines are results  obtained using Eq.(\ref{eq:ImprovedScaling}), while the dashed lines  correspond to Eq.~(\ref{eq:SimpleScaling}).}
\end{figure*}

\subsection{Transverse momentum spectra of light nuclei}
Figure~\ref{fig:fig1} presents our theoretical predictions for the invariant transverse momentum ($p_T$) spectra of protons, deuterons, and ${^3\mathrm{He}}$ in Pb+Pb collisions at $\sqrt{s_{NN}} = 5.36$ TeV. The spectra are  evaluated at midrapidity ($-0.5 < y < 0.5$)  for three centrality classes and exhibit several features characteristic of the coalescence mechanism and collective hydrodynamic expansion. First, the overall production yields show a strong centrality dependence. Second, the large radial flow developed during the QGP and hadronic evolution phases  induces a mass-dependent blue-shift to the spectra, efficiently boosting heavier  nuclei to higher transverse momenta. Third, a comparison of the spectral magnitudes across species in Fig.~\ref{fig:fig1} reveals a drastic suppression in yield  with each additional nucleon in the nuclei. This exponential decrease, spanning several orders of magnitude from protons to $^3$He, reflects the intrinsic coalescence penalty factor.  These predicted spectra can be directly compared with the upcoming ALICE measurements from LHC Run~3.
\par{}
\begin{figure*}[!t]
	\centering
	\includegraphics[width=0.9\textwidth]{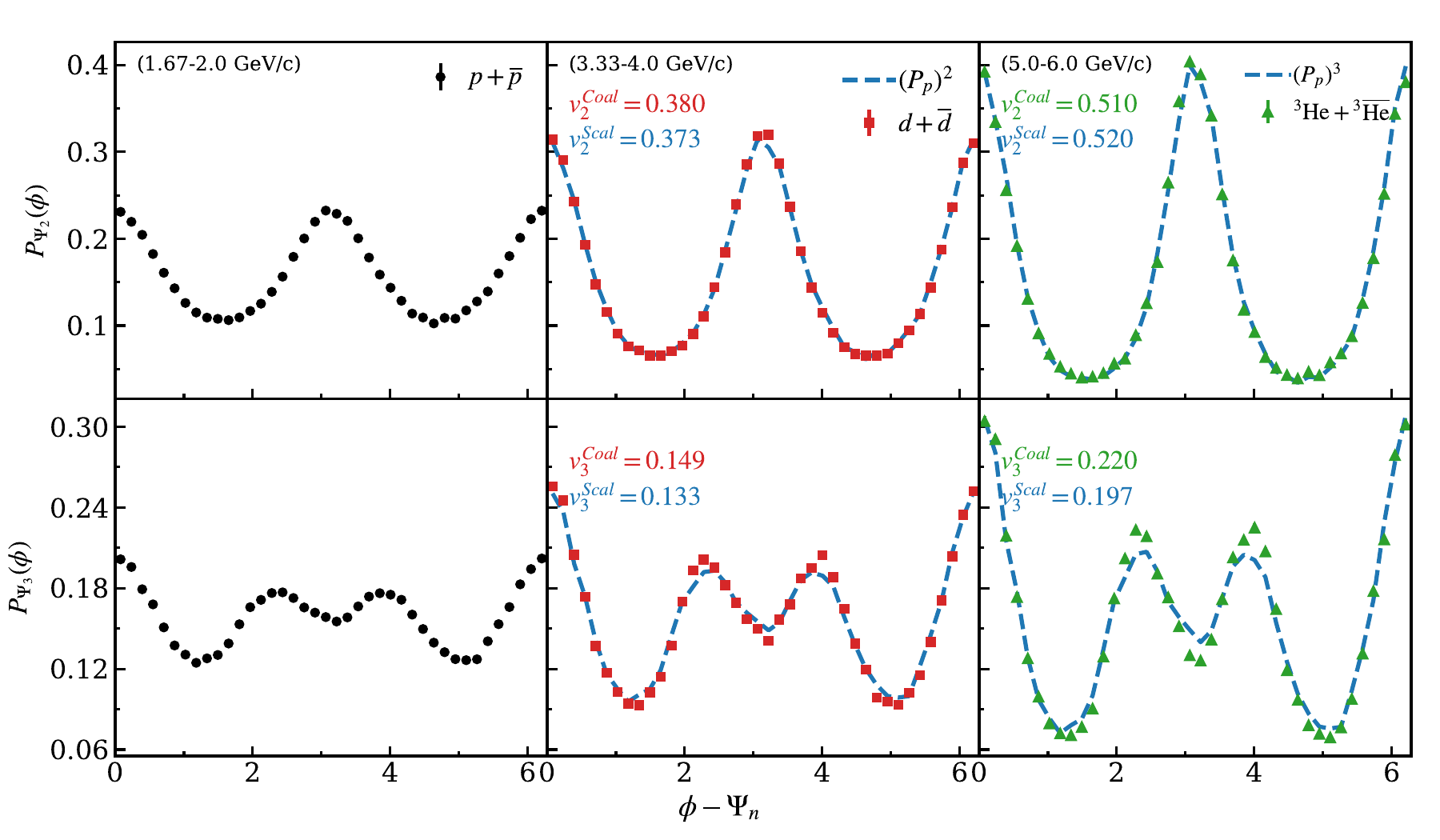}
    \caption{\label{fig:fig3}
    Normalized azimuthal distributions of proton (left), deuteron (middle) and $^3$He (right) relative to $\Psi_2$ (top) and $\Psi_3$ (bottom) in the 40\%–60\% centrality class. Symbols  represent the result from the coalescence model  ($v^{Coal}_n$), while  the dashed lines  show the scaling expectation  $\left(\frac{d N_p}{d\phi}\right)^{A}$ ($v^{Scal}_{n}$). The transverse momentum ranges are
    $1.67\!<\!p_T\!<\!2.0\,$GeV/$c$ (proton), $3.33\!<\!p_T\!<\!4.0\,$GeV/$c$ (deuteron),
    and $5.00\!<\!p_T\!<\!6.0\,$GeV/$c$ ($^3$He). The extracted flow coefficients are shown in each panel. 
    }
\end{figure*}

\subsection{\label{NCQ-scaling of vn}Nucleon number scaling  of light nuclei anisotropic flow}
Using the MUSIC+UrQMD+COAL model, we also calculate the $v_2$ and $v_3$ of protons, deuterons, $^3\mathrm{He}$, $^3_\Lambda\mathrm{H}$, and their  corresponding antiparticles  in Pb+Pb collisions at $\sqrt{s_{NN}}=5.36$~TeV. The anisotropic flow coefficients are determined via the event plane method,
\begin{equation}
    v_{n}= \langle \cos(n(\phi-\Psi_n)) \rangle /R_{n},
\end{equation}
where  $\phi$ is the azimuthal angle of the identified particle, $\Psi_n$ is the reconstructed $n$th-order event-plane angle, and $R_n$ is the corresponding event-plane resolution. The  average is taken over all particles of a given species and over all events within a specific centrality class. The event-plane resolution $R_n$ is calculated for each centrality bin using the standard three-subevent method~\cite{Poskanzer:1998yz}: 
\begin{equation}
    R_n = \sqrt{\frac{\langle \cos(n(\Psi^{A}_{n}-\Psi^{B}_{n}))\rangle \langle \cos(n(\Psi^{A}_{n}-\Psi^{C}_{n}))\rangle}{\langle \cos(n(\Psi^{B}_{n}-\Psi^{C}_{n}))\rangle}},
\end{equation}
where $\Psi^{A}_{n}$, $\Psi^{B}_{n}$ and $\Psi^{C}_{n}$ are the event plane angles reconstructed in three separate pseudorapidity ($\eta$) windows: $-3.7 < \eta_{A} < -1.7$ and $2.8 < \eta_{A} < 5.1$, $0.0 < \eta_{B} < 0.8$, and $-0.8 < \eta_{C} < 0.0$. The event-plane angle in each window is obtained from 
\begin{equation}
    \Psi_{n}=\frac{1}{n}\mathrm{arctan}\left(\frac{\text{Im}(Q_n)}{\text{Re}(Q_n)}\right),
\end{equation}
where the flow vector $Q_n$ is defined as 
\begin{equation}
    Q_n = \sum^{M}_{i}\mathrm{exp}(in\phi_{i}),
\end{equation}
with $M$ denoting the number of particles in the corresponding pseudorapidity window.

The upper panels of Fig.~\ref{fig:fig2}  show the $p_T$ dependence of $v_2$  for (anti-)proton (a), (anti-)deuteron (b), and (anti-)$^3$He (c) in the $0-20\%$, $20-40\%$, and $40-60\%$  centrality classes. The colored bands represent our results from the MUSIC+UrQMD+COAL model. The dashed lines  denote calculations based on the simple nucleon number scaling  relation in Eq.(\ref{eq:SimpleScaling}), while the solid lines denote results from the improved scaling relation in Eq.(\ref{eq:ImprovedScaling}). The simple scaling relation is seen to overestimate the $v_2$ of deuterons and $^3$He  at $p_T> 3$~GeV/$c$ in the 20\%-40\% and 40\%-60\% centrality classes, whereas the improved scaling  relation agrees well with the full model calculations up to $p_T\approx 8$ GeV/$c$. 

The lower panels of Fig.~\ref{fig:fig2} show the corresponding results for the $v_3$, obtained with the event-plane angle reconstructed using the three-subevents method. A non-zero (positive) $v_3$ is observed across the entire $p_T$ range, with a  notably weaker centrality dependence compared to that of $v_2$. The simple scaling relation yields larger $v_3$ values for deuterons and $^3$He than the improved scaling  relation; however, the differences remain small   due to the relatively small magnitude of the proton $v_3$. Consequently, both scaling relations agree surprisingly well with the full model calculations, shown as the colored bands.

\subsection{Angular distributions}
To further investigate the validity of the mass-number scaling of light nuclei anisotropic flow, we present in Fig.~\ref{fig:fig3} the normalized azimuthal angle distributions of protons, deuterons, and $^3$He. The results from the coalescence calculations using the MUSIC+UrQMD+COAL model (symbols) are compared with the scaling expectations obtained from the $A$-th power of the proton distribution (dashed lines). The transverse momentum ranges are selected to satisfy $p_T/A \approx \mathrm{const}$, ensuring a consistent kinematic comparison across species. The angular distributions of light nuclei are found to generally follow the shape of  $\left(\frac{d N_p}{d\phi}\right)^{A}$, confirming that light nuclei production is primarily governed by the collective flow of the constituent nucleons. However, a quantitative comparison of the extracted flow coefficients reveals subtle discrepancies between the coalescence results $v_n^{\mathrm{Coal}}$ and the scaling expectations $v_n^{\mathrm{Scal}}$. While $v_2^{\mathrm{Coal}}$ agrees reasonably well with $v_2^{\mathrm{Scal}}$, noticeable deviations are observed in the third-order flow, where $v_3^{\mathrm{Coal}}$ tends to differ from $v_3^{\mathrm{Scal}}$. This indicates that, although the simple scaling relation captures the dominant features of the angular distributions, it does not fully account for the higher momentum correlations and  higher-order flow fluctuations inherent in the coalescence process.
\begin{figure*}[!t]
	\centering
	\includegraphics[width=0.85\textwidth]{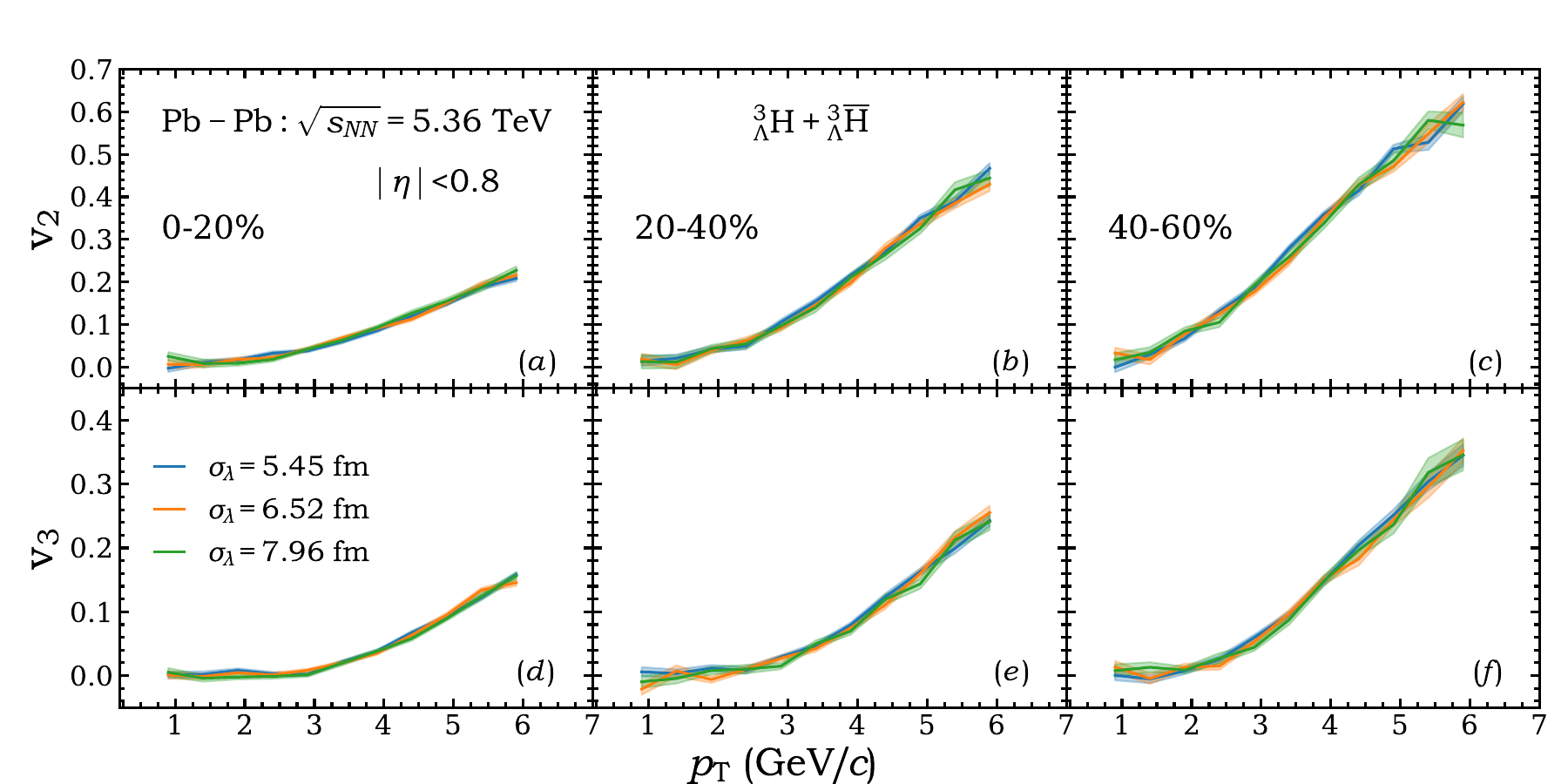}
    \caption{\label{fig:fig4}
          $v_2(p_T)$ and $v_3(p_T)$ of (anti-)hypertriton  for different values of the parameters $\sigma_{\lambda}$  in the $0-20\%$(left), 
        $20-40\%$(middle) and $40-60\%$(right) centrality classes in Pb+Pb collision at 
        $\sqrt{s_{NN}}=5.36$TeV. 
    }
\end{figure*}
\begin{figure*}[!t]
    \centering
    \includegraphics[width=0.85\textwidth]{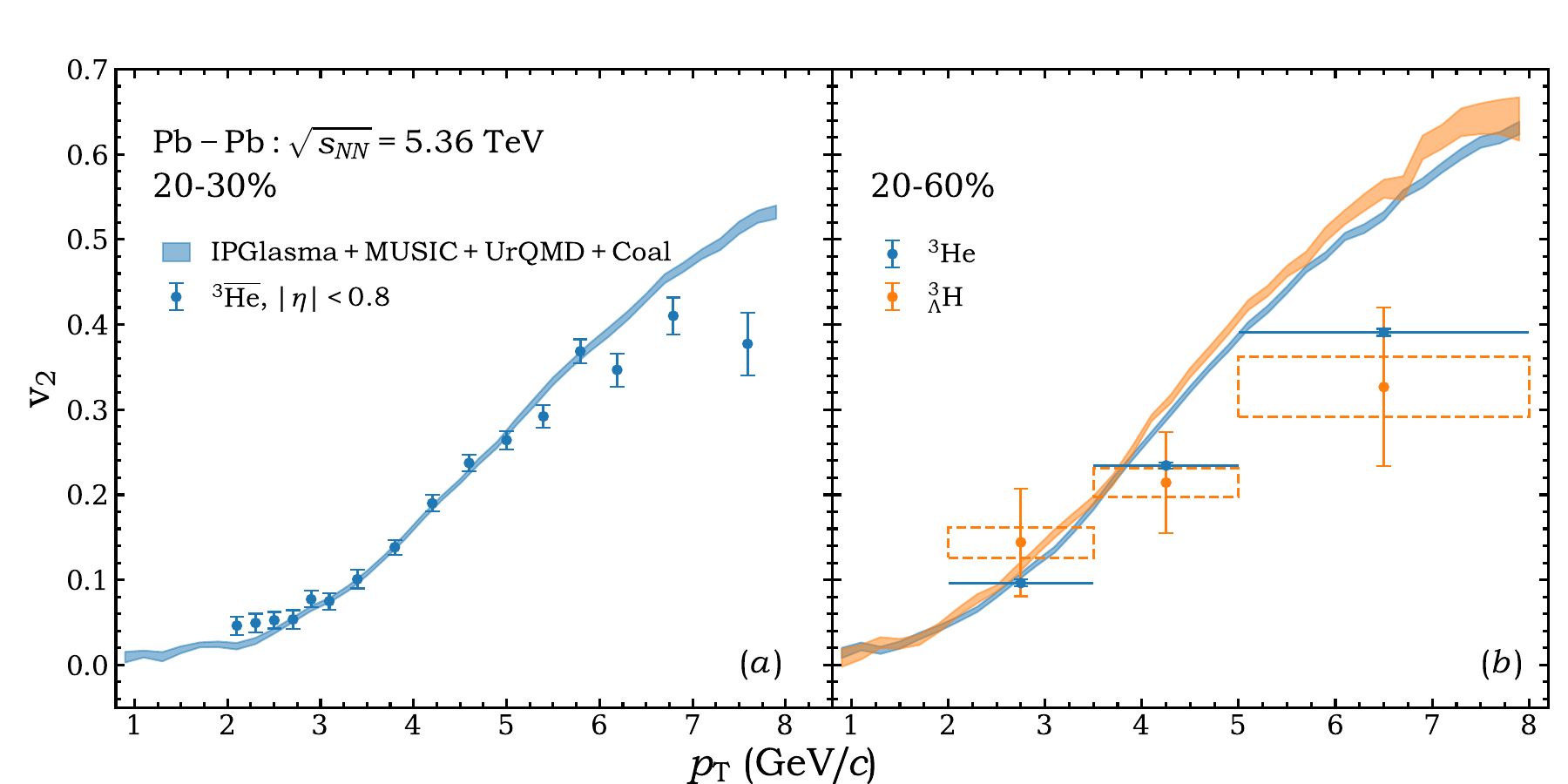}
    \caption{\label{fig:fig5}
     Panel (a): $v_{2}(p_T)$ of ($^3\overline{\mathrm{He}}$) in Pb–Pb collisions at $\sqrt{s_{NN}}=5.36\,$TeV for the 20–30\% centrality class.   Panel (b): $v_{2}(p_T)$ of $^3\mathrm{He}$ and $_{\Lambda}^3\mathrm{H}$ in the same system for the 20–60\% centrality
     class. The preliminary ALICE data  are taken from Ref.~\cite{alicecollaboration2026measurementellipticflow3he}.
    }
\end{figure*}
\subsection{\label{sec:level3B} Anisotropic flow of (anti-)hypertriton}
Figure~\ref{fig:fig4} shows the $p_T$    dependence  of $v_2$   (upper panels) and $v_3$ (lower panels)  of (anti-)hypertritons  for different centrality classes. The colored bands represent results for   three different $\Lambda-d$ separation distances, or equivalently, three different $\Lambda$ separation energies. Both the elliptic and triangular flow of the hypertriton are seen to increase from central to  peripheral collisions, with  magnitudes close to those of $^3$He. In contrast to its yield and $p_T$ spectrum~\cite{Liu:2024ygk}, its anisotropic flow coefficients show little sensitivity to the  $\Lambda-d$ separation distance.
\par{}

\subsection{Comparison with preliminary experimental data}
To assess the performance of the coalescence model in describing the collective flow of light (anti-)nuclei, we present in Fig.~\ref{fig:fig5} the transverse momentum dependence of the elliptic flow coefficient $v_{2}$ for $^3\overline{\mathrm{He}}$ in the $20$–$30\%$ centrality class (panel a) and for $^3\mathrm{He}$ and for $_{\Lambda}^3\mathrm{H}$ in the $20$–$60\%$ class (panel b). The shaded bands represent predictions from the IP-Glasma+MUSIC+UrQMD+coalescence framework, with the band widths reflecting the theoretical uncertainties. The markers denote the preliminary ALICE measurements~\cite{alicecollaboration2026measurementellipticflow3he}, where the vertical error bars represent statistical uncertainties for $^3\mathrm{He}$ and systematic uncertainties for $_{\Lambda}^3\mathrm{H}$, and the horizontal bars  indicate the $p_{T}$ bin widths. The theoretical predictions are found to be in excellent agreement with the experimental data across  both species and centrality intervals, demonstrating the ability of the nucleon coalescence mechanism  to capture the anisotropic flow of light nuclei. Future high-precision measurements will provide further  tests of this production mechanism.

\section{\label{sec:level4}Summary} 
In the present study, we have investigated the elliptic flow  and triangular flow   of (anti-)protons, (anti-)deuterons, (anti-)$^3$He, and (anti-)hypertriton ($^3_\Lambda$H) in Pb+Pb collisions at $\sqrt{s_{NN}} = 5.36$~TeV, employing a nucleon coalescence model with phase-space distributions of kinetically frozen-out nucleons generated by the hybrid IP-Glasma+MUSIC+UrQMD framework. We find that the simple scaling relation of $v_2$ with the number of constituent nucleons $A$ breaks down at high transverse momentum ($p_T/A > 1.5$~GeV/$c$), where it overestimates the $v_2$ of deuteron and $^3$He. An improved scaling relation, derived from the $A$-th power of the proton azimuthal distribution, remains valid up to $p_T/A \approx 3$~GeV/$c$ and provides good agreement with the full coalescence model calculations. For $v_3$, both scaling relations yield similar results owing to the relatively small magnitude of the proton $v_3$, and both agree well with the model calculations across all centrality classes considered.

We   also  present predictions for the $v_2$ and $v_3$ of the (anti-)hypertriton. Both flow coefficients are found to increase from central to peripheral collisions, with magnitudes close to those of $^3$He. Notably, unlike the hypertriton yield and $p_T$ spectrum, its anisotropic flow coefficients are insensitive to the $\Lambda$–$d$ separation distance within the hypertriton wave function. A comparison with preliminary ALICE measurements shows good agreement, demonstrating the ability of the coalescence model to describe the collective flow of light (hyper-)nuclei. These results, together with future high-precision data from LHC Run~3, will provide further insight into the production mechanisms of light (anti-)(hyper-)nuclei in high-energy heavy-ion collisions.

\section{Acknowledgments}
We thank Luca Barioglio, Chiara Pinto, and  Sourav Kundu for helpful discussions.  This work was supported in part by the National Key Research and Development Project of China under Grant No. 2024YFA1612500,  and by the National Natural Science Foundation of China under Contracts No. 12422509 and No. 12375121, and 12547102, and  by the Science and Technology Commission of Shanghai Municipality under Grant No. 23590780100. The computations in this research were performed using the CFFF platform of Fudan University.

%

\end{document}